# Instrument-Independent Dastgah Recognition of Iranian Classical Music Using AzarNet

Shahla Rezezadeh Azar
Computer Engineering Department
K.N. Toosi University of Technology
Tehran, Iran
ShahlaRezazadeh@email.kntu.ac.ir

Saber Malekzadeh
Computer Science Department
University of Tabriz
Tabriz, Iran
Saber.Malekzadeh@sru.ac.ir

Ali Ahmadi
Computer Engineering Department
K.N. Toosi University of Technology
Tehran, Iran
Ahmadi@kntu.ac.ir

Maryam Samami
Computer Engineering Department
Islamic Azad University, Sari branch
Sari, Iran
maryamsamami2013@gmail.com

*Abstract*— **In this paper, AzarNet, a deep neural network (DNN), is proposed to recognizing seven different Dastgahs of Iranian classical music in Maryam Iranian classical music (MICM) dataset. Over the last years, there has been remarkable interest in employing feature learning and DNNs which lead to decreasing the required engineering effort. DNNs have shown better performance in many classification tasks such as audio signal classification compares to shallow processing architectures. Despite image data, audio data need some preprocessing steps to extract spectra and temporal features. Some transformations like Short-Time Fourier Transform (STFT) have been used in the state of art researches to transform audio signals from time-domain to time-frequency domain to extract both temporal and spectra features. In this research, the STFT output results which are extracted features are given to AzarNet for learning and classification processes. It is worth noting that, the mentioned dataset contains music tracks composed with two instruments (violin and straw). The overall f1 score of AzarNet on test set, for average of all seven classes was 86.21% which is the best result ever reported in Dastgah classification according to our best knowledge.**

### Introduction

Music tags indicate the high-level information about a music sample, such as emotion (sadness, anger, or happiness), genre (Classical, Jazz, Pop) and also Dastgah [1] (Segah, Mahour, Shour). To predict music tags, a classification process can be used applying the audio signal [1].

Deep convolutional neural networks (DCNNs) have been applied extensively for various music classification tasks such as music tagging [1] [2], genre classification [3] [4]. Convolutional filters extract spectra feature in convolutional layers of Convolutional neural network (CNN) which is trained in order to classify unknown data [5] [6]. Extracting spectra features helps to estimate the tags, aiming to predict single or multi-label classification [1].

Lately, deep neural networks (DNNs) are provided which are capable to learn the relevant features automatically. DNNs have been applied extensively in audio analysis, speech recognition [7] and also auto-tagging [1] [8]. DNNs composed of multiple hidden layers with hidden units trained to represent some main structure of data [2].

In this paper, the Iranian Classical music samples are classified according to their Dastgahs. A DNN is proposed to extract time frequency domain features. The raw data is transformed from time domain to time frequency domain with Short-time Fourier transform (STFT). Finally, the improving of the gained f result will be demonstrated by applying DNN with convolution layers.

## I. RELATED WORK

In some papers, several DNN-based algorithms have been used for recognition of the class label of the music. In order to extract features from the audio signals in [9], Multi-resolution spectrograms are employed. To predict the class label of unseen dataset, the pre-trained weights of multilayer perceptron are being used in [8]. Two convolutional layers are provided in the proposed network in [6] which takes Mel-spectrograms and also raw audio signals as input features [1].

CNNs have been mixed with recurrent neural networks (RNNs) which is presented as a hybrid model to model sequential data. The hybrid model namely a convolutional recurrent neural network (CRNN) was first provided for document classification [10]. Thereafter, the given model used to image classification [11], music transcription [12] and music tagging [6].

In [11] RNNs are employed to aggregate the temporal patterns, on the other hand, CNNs take input for extracting local feature. After sub-sampling in convolutional and max-pooling layers, a feature map will be the output which is fed into the two-layer RNN that are located in the last hidden state, connected to the output of the network.

---
[1]- Dastgah is a musical modal system in Iranian classical music.

In [13] sample-level DCNN models are presented which take raw waveforms as input and is able to learn representations from very small grains of waveforms. In the proposed model, the strides of the first convolutional layer decline from frame-level to sample-level. The deep structure with sample-level filters improve the accuracy in music auto-tagging. To recognizing hierarchically learned features which are sensitive to log-scaled frequency along layer, such as Mel-frequency spectrogram, the filters trained in a sample level DCNN in each layer.

The new provided approaches are based on mid-level representations of music, e.g. spectrograms despite the older ones which employ extract features from raw audio signals. In [12] a CNN is trained using feature extracting on spectrograms as well as raw audio signals directly. The observed performances on the two given approaches on automatic tagging illustrate that, spectrogram-based approach outperforms comparing to the raw audio signals approach.

## II. FUNDAMENTAL CONCEPTS

In this section, at first, the applied dataset is described. Then the performance and structure of STFT and DNN are represented respectively to represent their better performance in music classification.

### A. Maryam Iranian Classical music dataset (MICM)

Maryam Iranian Classical music dataset is proposed in this paper for the very first time. The given dataset includes 1137 music samples which contains 631 music samples with the foreground straw instrument sound and also some other instruments in the background. The remained music samples include the violin instrument sound as the foreground instrument sound. Aiming to present an instrument Independent method to classifying Dastgahs, two musical instruments are employed, Violin and straw. The MICM dataset contains seven classes which provides several music samples in seven Iranian Classical Dastgahs namely as follows: Shour, Homayoun Mahour, Segah, Chahargah, Rastpanjgah and Nava. Each music samples have different numbers of signal samples and also the sample rate of each music sample is 8192.

The Table 1 illustrates the numbers of music samples in each class.

TABLE I. MICM SAMPLES DESCRIPTION

| Name of Dastgah | number of music samples |
|---|---|
| Shour | 445 |
| Homayoun | 173 |
| Mahour | 150 |
| Segah | 74 |
| Chahargah | 106 |
| Rastpanjgah | 94 |
| Nava | 95 |

### B. Preprocesing of MICM music samples

As known, DNNs just are able to get data samples with the same length as inputs. As mentioned in the previous sub-section, each sound sample included in the MICM has different lengths. Each sound sample is cut. The all cut music samples have 131072 signal samples. As previously mentioned the sample rate of each music sample is 8192. Therefore, each sound sample contains 16 seconds of music. The reason behind this selection is because, Dastgahs in Iranian Classical music can be recognized easily with 16 seconds of music.

### C. Short-time Fourier transform (STFT)

The Short-time Fourier transform has applied extensively to preprocessing step for extracting the frequency features of the audio signal. To overcome the disadvantages of Fourier analysis, such as not having the ability to reflect the local time-domain information, Short-Time Fourier Transform (STFT) are used in this paper to represent the information of the signals. At first, STFT divides the signal into small time blocks; secondly, it uses the Fourier transform to each block [14].

$$F(\omega) = \int_{-\infty}^{+\infty} f(t) \exp(-i\omega t)\, dt \ . \quad (1)$$

Where $I = \sqrt{-1}$.

The STFT formula for the time domain signal F(t) is as follows:

$$F_{STFT}(\tau, \omega) = \int_{-\infty}^{+\infty} f(t)\, g^*(t-\tau) \exp(-i\omega t) dt \ . \quad (2)$$

Where $\tau$ is the time shift parameter, the signal g(t) is a fixed length window and the symbol (∗) expresses the complex conjugate [15].

### D. DNN

A DNN is a feed-forward DNN that usually composed of more than one hidden layer of neurons between its inputs and outputs. Each hidden neuron, j, employs the logistic function to map the received input from the previous layer, $x_j$, to the scalar state, $y_j$, as its output which is sent to the next layer.

$$y_j = \text{logistic}(x_j) = \frac{1}{1+e^{-x_j}}, \quad x_j = b_j + \sum_i y_i w_{ij} \ . \quad (3)$$

Where $b_j$ is the bias of neuron j, i is an index for the all neurons in the previous layer, and to connecting between neuron i and neuron j in the next layer, there is a connection and also a weight on it namely $w_{ij}$. In order to classifying multi class dataset, neuron j change its input into a class probability.

For multiclass classification, output of neuron j should be a class probability. Neuron j employs the "softmax[2]" non-linearity function to change the total input of neuron j, $x_j$, into a class probability as an output.

$$p_j = \frac{\exp(x_j)}{\sum_k \exp(x_k)} \ . \quad (4)$$

where k is an index over all classes.

DNNs are trained by back propagating derivatives of the used cost function that compares the discrepancy among the predicted outputs and the ground truth labels [15].
In the proposed model, softmax and cross-entropy are used as an output function and a cross-entropy respectively. The cost function C computes the cross-entropy between target probabilities, d and the outputs of the softmax, p:

$$C = -\sum_j d_j \; log p_j. \qquad (5)$$

CNNs are one particular type of deep, feed forward network composed of kernels that have learnable weights. Each kernel convolves on an input data and activation function is applied to the given convolution result. A CNN is a kind of score function as receives a STFT sample on one end to output class scores at the other end. CNNs also contain a loss function to calculate the cost of the network prediction to be idol and optimizing results by reforming the weights in back propagation operation with an optimizer function [16].

In the proposed deep model, Gated Recurrent unit (GRU) a new kind of RNN layer is used. A RNN is a type of artificial neural network where connections among neurons create a directed graph along a sequence. RNNs employ their memory to process sequences of inputs [17]. RNN, relates all the sequences of inputs together. In the prediction or generation cases, the relation among all the previous words or samples helps in predicting or generating the better result. The RNN produces the networks with loops in them, causing to persist the information [18].

### III. THE PROPOSED METHOD

The architecture of the proposed model AzarNet, is a kind of DNN in which some several layers are applied to extract and learn spectra and temporal features from STFT. The raw data which is a representation of temporal features of music samples is transformed from time domain to time-frequency domain with STFT. The given STFT features are used as input of the first layer of AzarNet. AzarNet architecture is illustrated in detail.

#### A. Preprocesing of MICM music samples

As known, DNNs just are able to get data samples with the same length as inputs. As mentioned in the previous sub-section, each sound sample included in the MICM has different lengths. Each sound sample is cut. The all cut music samples have 131072 signal samples. As previously mentioned the sample rate of each music sample is 8192. Therefore, each sound sample contains 16 seconds of music. The reason behind this selection is because, Dastgahs in Iranian Classical music can be recognized easily with 16 seconds of music.

#### B. STFT application on the preprocessed data

STFT is applied to the given preprocessed data which is in time-domain to have a time-frequency domain output. While MICM dataset contains music samples without noise STFT will be a better choice in comparison with similar functions like mel-frequency Cepstrum coefficients (MFCCs). This function includes some input parameters which affect the result of the function such as changing the size and resolution of the output.

One of these parameters is fast Fourier transform (FFT) windows size which is set to 510. The second parameter is hop length that is the number of the frames of audio between STFT columns. This parameter is set to 514. By using the mentioned input parameters and giving a preprocessed audio signal as input signal, the output size will be 256*256 matrix that shows both spectra and temporal features of the input audio signal in time-frequency domain. The figure 1 shows sample STFT output. All amplitudes in STFT results are scaled between -80 and 0 and because in DNN (especially in relu activation function) zero and negative numbers are worthless, so +80 is added to all numbers in STFT (Max value was 80 and min value was 0.).

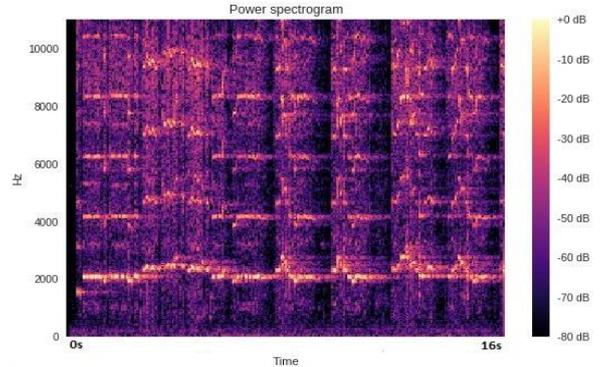

Figure 1.     The scaled STFT sample.

#### C. Deep learning

Deep learning has employed extensively as a learning technique in DNNs which shows a great performance in the most of classification tasks.

In this paper, AzarNet, a DNN which is composed of convolution, GRU and fully connected layers is proposed. The network is starting by a convolution layer as shown in figure 2. As we go through the layers of AzarNet, the depth of outputs will be enhanced by increasing the numbers of filters in each convolutional layer gradually. DNNs are able to learn hierarchical features. Convolution layers are used to extract spectra features. After flattening the output of convolution layers, the results are fed to the two GRU layers. Using recurrent layers to recognize the patterns of time series, outperforms compared to CNN, since RNNs have the ability to aggregating the temporal pattern. RNNs are usually used to recognize the Dastgah patterns in this paper    these temporal features, due to temporal features-based Dastgah patterns.

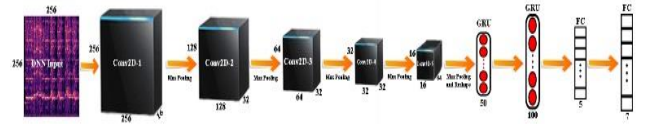

Figure 2.     AzarNet achitecture

The output of the last convolution layer which is sent to the reshape layer, is ordered by time just like the STFT samples given to the first layer of the network. The temporal feature which is the output of the reshape layer, fed into the first GRU layer. The first GRU layer extracts the temporal feature and returns a 2d sequence of spectra features which are ordered by time and sends them to the next GRU layer. The reason for using the GRU layers after several max pooling layers is related to the weakness of recurrent layers which is limited memory. The mentioned disadvantage makes recurrent layer unable to learn all patterns of a big data sample. The output of the second GRU layer is a 1d feature matrix that is not ordered by time or frequency, since it is given to a fully connected layer.

The first fully connected layer is a bottleneck with only five neurons. This kind of layers have shown good results in extracting special features commonly named as bottleneck features. Bottleneck features are very useful in recognizing acoustic data in previous researches [19]. The number of neurons included in the last layer which is another fully connected layer are equal to the number of classes. It is worth noting that, class probability is the last output which is computed by "softmax" non-linearity function that commonly has used in order to classify multiclass data set. This type of layer is commonly known as a classifier layer.

Each convolutional layer contains a max pooling layer with 2*2 pooling size and the stride of 2*2 to reduce the size of the output of the convolution layer. These reductions help the network to extract bigger and bigger spectra curves from the convolution output during the training process. So as all kernels of all convolution layers are 3*3.

The first convolution layer extracts small spectra curves in the 2d input while the last convolution layer extracts big and general features. Worth to mention that all convolution layers has same padding to keep the output size same as the input size.

For better regularization in data, four batch normalization layers are used with the momentum of 0.8. These types of layers can have positive effects on learning process of the DNNs such as making activation functions viable, providing regularization to avoid overfitting, increasing speed of the training process and making weights initialization much easier in deep networks [20].

Also for more regularization to avoid overfitting 5 dropout layers are used between convolutional layers. Their dropout rates are respectively 0.1, 0.2, 0.3, 0.3 and 0.4. Since the number of parameters (convolution parameters) are increased during the network as shown in Table 2, increase in dropout rate improved regularization [21].

Both kernel and activity regularizes are used in convolutional and GRU layers. Both of them are a combination of least absolute deviations (LAD) and least squares error (LSE) with 0.01 penalties on layer parameters and activities during optimization. The reason for using the combination of these functions is because the combination structure is more robust and always has multiple stable solutions, however, each of the given functions does not have the mentioned advantages individually [22].

As an activation function for all layers except the last layer (classifier which uses softmax) Leakyrelu with an alpha value of 0.1 is used due to the given experimental results [23].

A brief diagram of layers and parameters of AzarNet is shown in Table 2.

TABLE II. DIAGRAM OF LAYERS OF AZARNET

| Layer type | Output shape | # Parameters |
|---|---|---|
| 2D Convolution (3*3)(16) | (256, 256, 16) | 160 |
| Dropout (0.1) | (256, 256, 16) | 0 |
| Batch Normalization (0.8) | (256, 256, 16) | 64 |
| 2D Max Pooling (2*2) | (128, 128, 16) | 0 |
| 2D Convolution (3*3)(32) | (128, 128, 32) | 4640 |
| Dropout (0.2) | (128, 128, 32) | 0 |
| Batch Normalization (0.8) | (128, 128, 32) | 128 |
| 2D Max Pooling (2*2) | (64, 64, 32) | 0 |
| 2D Convolution (3*3)(32) | (64, 64, 32) | 9248 |
| Dropout (0.3) | (64, 64, 32) | 0 |
| Batch Normalization (0.8) | (64, 64, 32) | 128 |
| 2D Max Pooling (2*2) | (32, 32, 32) | 0 |
| 2D Convolution (3*3)(32) | (32, 32, 32) | 9248 |
| Dropout (0.3) | (32, 32, 32) | 0 |
| Batch Normalization (0.8) | (32, 32, 32) | 128 |
| 2D Max Pooling (2*2) | (16, 16, 32) | 0 |
| 2D Convolution (3*3)(64) | (16, 16, 64) | 18496 |
| Dropout (0.4) | (16, 16, 64) | 0 |
| Batch Normalization (0.8) | (16, 16, 64) | 256 |
| 2D Max Pooling (2*2) | (8, 8, 64) | 0 |
| Reshape | (64, 64) | 0 |
| GRU (50) | (64, 50) | 17400 |
| GRU (100) | (100) | 45600 |
| FC (5) | (5) | 505 |
| FC (7) (classifier) | (7) | 42 |

IV. RESULTS

After training the AzarNet on 80% of data mentioned in Table 1, the network tested on 20% of the unknown data (equally 20% of each class). The results are also compared with and without using GRU and bottleneck layers.

The overall result of using convolutional layers without any GRU and bottleneck layer was 84.80%. Detailed results are shown in Table 3.

TABLE III. DNN RESULTS WITHOUT GRU AND BOTTELNECH

| Name of Dastgah | Precision | Recall | F1 |
|---|---|---|---|
| Shour | 96.35 | 85.94 | 90.85 |
| Homayoun | 75.34 | 81.28 | 78.20 |
| Mahour | 90.42 | 87.12 | 88.74 |
| Segah | 81.79 | 86.16 | 83.92 |
| Chahargah | 62.23 | 86.57 | 72.41 |
| Rastpanjgah | 88.06 | 90.74 | 89.38 |
| Nava | 94.57 | 86.10 | 90.14 |

The overall result of AzarNet was 86.21% detailed results are also shown in Table 4.

TABLE IV. MAIN OVERALL RESULTS

| Name of Dastgah | Precision | Recall | F1 |
|---|---|---|---|
| Shour | 97.42 | 87.52 | 92.21 |
| Homayoun | 76.22 | 82.72 | 79.34 |
| Mahour | 91.53 | 89.92 | 90.72 |
| Segah | 83.58 | 84.95 | 84.26 |
| Chahargah | 63.04 | 91.53 | 74.66 |
| Rastpanjgah | 90.66 | 90.30 | 90.48 |
| Nava | 96.12 | 87.92 | 91.84 |

The average of F1 scores as seen in Table 3 is 86.21, which is the best result has ever reported in case of classification of seven classes MICM data set.

The best state of art results are from [24] and [25]. In [24] a dataset contains seven classes which provided seven several Dastgahs of Iranian classical music are classified with one layer artificial neural network on FFT samples. The f1 classification result was 83%. Also in [25] a dataset contains seven classes which provided seven Dastgahs of Iranian classical music are classified with one layer artificial neural network on FFT samples. The best classification accuracy reported in this paper is 72%.

## V. CONCLUSION

In this paper AzarNet is proposed to classify a data set which contains 1137 violin and straw sounds with 7 classes. Each class label represents seven several different Dastgahs which belongs to Iranian classical music. the data set is applied to train and test the proposed network is named MICM music data set that is used for the first time. STFT is used to extract both spectral and temporal features of raw music data. Then STFT samples are given to AzarNet for the classification process. The overall result of classifying on 20% of the entire data set, known as test data set, was 86.21% which is the best result has ever reported. Worth to note that this result can be improved with more high quality data and more complex neural networks in future.